\documentclass[11pt,a4paper]{article}

\usepackage[T1]{fontenc}
\usepackage[utf8]{inputenc}
\usepackage{lmodern}

\usepackage{graphicx}
\usepackage{amsmath,amssymb}
\usepackage{ragged2e}
\usepackage{geometry}
\justifying

\begin{document}

\title{Beyond Hexagonal Boron Nitride: First-Principles Study of Pentaoctite-BN and Pop-BN Monolayers}

\author{
Victor M. S. da Conceição$^{1}$,
Erika N. Lima$^{2}$,
Roberto H. Miwa$^{1}$,
Igor S. S. de Oliveira$^{3}$\\[0.5em]
{\small $^{1}$Instituto de Física, Universidade Federal de Uberlândia, Uberlândia, 38400-902, MG, Brazil}\\
{\small $^{2}$Instituto de Física, Universidade Federal de Mato Grosso, Cuiabá, 78060-900, MT, Brazil}\\
{\small $^{3}$Departamento de Física, Universidade Federal de Lavras, Lavras, 37203-202, MG, Brazil}\\[0.5em]
{\small E-mail: manoelsoares1652@gmail.com}
}

\date{}

\maketitle

\begin{abstract}
\justifying
We investigate two novel non-hexagonal boron nitride monolayers, pentaoctite-BN (PO-BN) and pop-BN (PP-BN), using first-principles calculations. Their structural, electronic, mechanical, vibrational, thermal, and optical properties are systematically analyzed to assess their stability and potential applications. Despite being metastable with respect to hexagonal BN, both polymorphs satisfy the criteria for dynamical, mechanical, and thermal stability, indicating that they are viable two-dimensional materials. Both systems are indirect-gap semiconductors whose electronic states near the band edges are dominated by out-of-plane $p_z$ orbitals. Their distinct pentagon–octagon ring networks also give rise to different in-plane elastic anisotropies. Many-body optical calculations reveal strong excitonic effects and pronounced polarization-dependent optical absorption, with lattice engineering shifting the optical response from the ultraviolet toward the visible and infrared regions. These findings demonstrate that engineering non-hexagonal lattice architectures provides an effective strategy for tuning the electronic and optical properties of two-dimensional BN, highlighting PO-BN and PP-BN as promising candidates for future optoelectronic and photonic applications.
\end{abstract}

\section{Introduction}

Two-dimensional (2D) materials have become a central platform for exploring the interplay between lattice architecture and physical properties. Among them, hexagonal boron nitride (\textit{h}-BN) occupies a distinctive position, it is isostructural and nearly lattice-matched to graphene, yet it is a wide-band-gap insulator, with an indirect gap of 5.96~eV~\cite{Cassabois2016}, and it has been available as an exfoliated monolayer since the first isolation of 2D atomic crystals~\cite{Novoselov2005}. These characteristics underpin its use as a substrate, encapsulant, and tunnel barrier in van der Waals heterostructures, as well as a host for single-photon emitters~\cite{im2025quantum,sajid2020single,yankowitz2019van}.

In parallel, the search for planar lattices beyond the honeycomb arrangement has led to several carbon allotropes composed of non-hexagonal rings. One example is penta-graphene, which is built entirely from pentagonal rings~\cite{Zhang2015}. Another is the family of two-dimensional carbon lattices based on the 5--8--5 ring network, first identified as the OPG-L and OPG-Z phases~\cite{Su2013}, later known as popgraphene and pentaoctite carbon, respectively. Popgraphene was subsequently investigated for energy-storage applications~\cite{Wang2018}, and the same pentagon--octagon ring network has also been predicted in silicon as popsilicene~\cite{Lima2024}. The synthetic relevance of non-hexagonal 2D lattices was reinforced by the bottom-up synthesis of the biphenylene network, a periodic $sp^2$ carbon lattice containing four-, six-, and eight-membered rings, obtained through an on-surface dehydrofluorination reaction between adjacent polymer chains~\cite{Fan2021}. In addition, extended 5--8--5 line defects have been realized in graphene with atomic precision~\cite{Lahiri2010}, indicating a possible route toward lattices assembled from pentagon--octagon building blocks. These advances demonstrate that periodic non-hexagonal ring networks are not merely theoretical constructions, but realistic targets for synthesis.

The pentaoctite lattice was later predicted beyond carbon in two-dimensional bismuth. In this system, the structure originates from the periodic repetition of an extended 5--8 line defect in hexagonal bismuthene, where the one-dimensional defect hosts topologically protected metallic states~\cite{Lima2016}. The corresponding periodic lattice was subsequently predicted to behave as a strain-tunable topological insulator~\cite{Lima2019}. This structural family was later extended to the group-V elements P, As, and Sb~\cite{daRosa2021}, and to the group-IV elements C, Si, Ge, and Sn~\cite{Kegler2025}. Its recurrence across elemental families, ranging from covalent $sp^2$ carbon to heavy spin--orbit-coupled bismuth, indicates that the pentagon--octagon ring network constitutes a robust structural motif rather than a peculiarity of a single chemical system.

Although the family of non-hexagonal carbon allotropes has expanded rapidly, the corresponding set of two-dimensional BN polymorphs remains comparatively sparse~\cite{Shahrokhi2017,REF_IrBN}. This limitation has a chemical origin. In a binary B--N lattice, complete B/N alternation can be maintained only in even-membered rings. Consequently, any odd-membered ring necessarily introduces homonuclear B--B or N--N bonds, which are energetically less favorable than heteropolar B--N bonds. The same preference is observed in BN fullerenes, which form square--hexagon cages rather than pentagon--hexagon ones~\cite{Jensen1993}, and in dislocation cores of \textit{h}-BN, where homonuclear-bond-free 4$|$8 pairs are favored over the 5$|$7 pairs commonly found in graphene~\cite{Liu2012}. The few reported non-hexagonal BN lattices therefore pay this energetic cost. For example, the three Irida-B$_{12}$N$_{12}$ phases contain triangular rings that inevitably introduce B--B and N--N bonds and were predicted to be wide-gap semiconductors, with HSE06 band gaps ranging from 2.43 to 3.20~eV~\cite{REF_IrBN}. Since both the pentaoctite and pop lattice architectures also contain pentagonal rings, their BN counterparts provide an ideal platform for assessing whether non-hexagonal BN networks can remain structurally stable and semiconducting despite the presence of homonuclear bonds.

In this work, we employ first-principles calculations to investigate two boron nitride polymorphs based on the pentaoctite and pop lattice architectures, denoted as PO-BN and PP-BN. We assess their energetic, dynamical, thermal, and mechanical stability and characterize their structural, electronic, vibrational, elastic, and optical properties. By comparing these non-hexagonal BN polymorphs with monolayer \textit{h}-BN, we elucidate how the underlying ring network influences the physical properties of two-dimensional BN, providing insight into the opportunities and limitations of lattice engineering in binary 2D materials.

\section{Computational Methods}

First-principles calculations were performed within the framework of Density Functional Theory (DFT), as implemented in the Vienna \textit{Ab initio} Simulation Package (VASP)~\cite{Kresse1996a,Kresse1996b}. The interaction between the valence electrons and the ionic cores was described using the projector-augmented wave (PAW) method~\cite{Kresse1999}. The exchange--correlation energy was treated within the generalized gradient approximation (GGA), employing the Perdew--Burke--Ernzerhof (PBE) functional~\cite{Perdew1996}. In addition, the screened hybrid HSE06 functional~\cite{Heyd2003,Krukau2006} was used to obtain a more accurate description of the electronic band structures, with the standard mixing parameter $\alpha = 0.25$ for the short-range Hartree--Fock exchange fraction and screening parameter $\mu = 0.2$~\AA$^{-1}$.

The structural optimization and electronic properties of the PO-BN and PP-BN monolayers were investigated using a plane-wave kinetic energy cutoff of 550~eV. The Brillouin zone was sampled using $\Gamma$-centered Monkhorst--Pack meshes~\cite{Monkhorst1976} of $7\times10\times1$ and $7\times5\times1$ $k$-points for PO-BN and PP-BN, respectively; this choice compensates for the distinct in-plane aspect ratios of the two unit cells ($a/b \approx 1.37$ for PO-BN and $a/b \approx 2.45$ for PP-BN), yielding an approximately uniform $k$-point density along the reciprocal-lattice directions of both structures. All lattice parameters and atomic positions were fully relaxed until the total energy difference between consecutive electronic steps was smaller than $10^{-6}$~eV and the residual atomic forces were below $2\times10^{-2}$~eV/\AA. A vacuum region of 20~\AA\ was included perpendicular to the monolayer plane to avoid artificial interactions between periodic images.

The dynamical stability of the structures was analyzed through phonon dispersion calculations using the finite-displacement method, as implemented in the Phonopy package~\cite{Togo2023,Togo2023b}. For these calculations, $4\times4\times1$ supercells were constructed from the optimized unit cells. A plane-wave cutoff energy of 550~eV and a $2\times2\times1$ $k$-point mesh were employed. A stricter electronic convergence criterion of $10^{-9}$~eV was adopted to ensure accurate force constants.

The mechanical properties were investigated by calculating the elastic constants using the energy--strain approach. The deformed configurations were generated using the VASPKIT package~\cite{VASPKIT}, while the total energies and stress tensors were obtained from VASP calculations. The calculated elastic constants were used to evaluate the mechanical stability and the anisotropic mechanical response of the PO-BN and PP-BN monolayers.

The thermal stability at finite temperature was examined using \textit{ab initio} molecular dynamics (AIMD) simulations within the canonical (NVT) ensemble. The simulations were performed at 300~K using the Nos\'e--Hoover thermostat~\cite{Nose1984,Hoover1985}, with a time step of 1~fs. The systems were evolved for 10~ps, and the time evolution of the total energy, temperature, and atomic configurations was monitored to verify the preservation of the pentagon--octagon lattices.

The optical properties were investigated using the GPAW code~\cite{Enkovaara2010,Mortensen2005}. Starting from the optimized structures obtained with VASP, ground-state calculations were performed within the PBE approximation using a kinetic energy cutoff of 500~eV. The Brillouin zone was sampled using $\Gamma$-centered Monkhorst--Pack meshes of $10\times12\times1$ and $10\times13\times1$ $k$-points for the PO-BN and PP-BN monolayers, respectively. A vacuum region of 20~\AA\ was maintained along the out-of-plane direction. To obtain a sufficient number of unoccupied electronic states for the excited-state calculations, a fixed-density calculation was performed, including a total number of bands equal to three times the number of occupied bands.
The optical response was calculated within the random-phase approximation (RPA) \cite{Yan2011} and by solving the Bethe--Salpeter equation (BSE) \cite{Onida2002}, as implemented in the GPAW response module. To account for the underestimation of the band gap at the semilocal DFT level, a rigid scissor correction was applied to the conduction bands so as to reproduce the HSE06 band gaps. Consequently, the calculated spectra are hereafter referred to as HSE06+RPA and HSE06+BSE. The dielectric matrix was expanded using a plane-wave cutoff energy of 90 eV for the response function. The BSE Hamiltonian was constructed including 9 occupied (valence) and 9 unoccupied (conduction) bands, and the Tamm--Dancoff approximation (TDA) \cite{Sander2015} was employed throughout the calculations. The complex dielectric function, $\varepsilon(\omega)$, was evaluated for light polarization along the $x$ and $y$.

\section{Results and Discussion}

\subsection{Structural Properties and Energetic Stability}

\begin{figure}[!htb]
 \centering
        \includegraphics[width=0.7\textwidth]{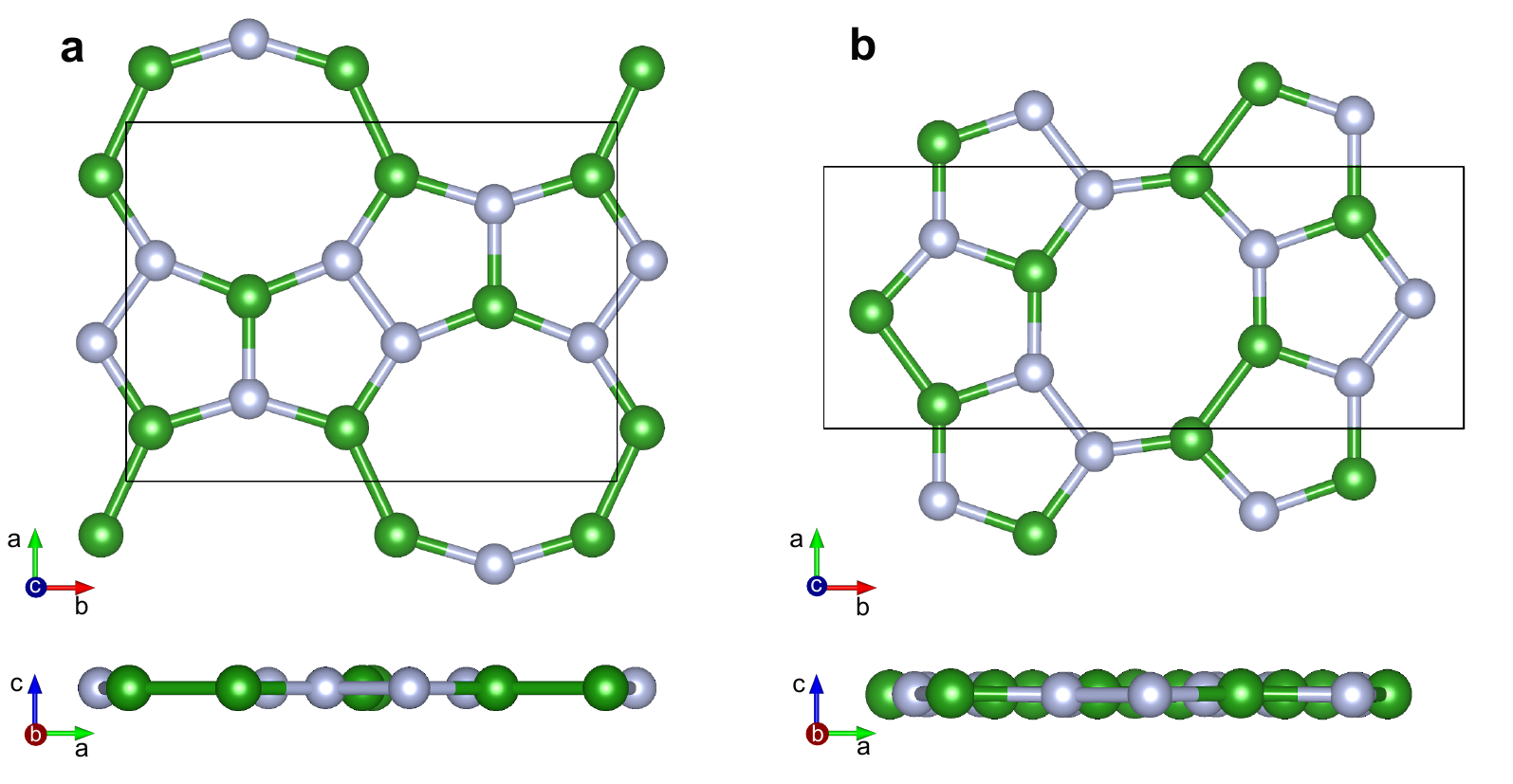}
 \caption{Optimized atomic structures of (a) PO-BN and (b) PP-BN monolayers, shown in top view (upper panels, unit cell outlined) and side view (lower panels). Green and silver spheres denote B and N atoms, respectively. Homonuclear B–B and N–N bonds, required to close the pentagonal rings, are visible as same-colored bonded pairs.}
\label{fig1}
\end{figure}

Figure~\ref{fig1} shows the optimized atomic structures of the two BN-based 2D lattices investigated in this work, denoted as pentaoctite-BN (PO-BN) and pop-BN (PP-BN). Both structures are composed of pentagonal and octagonal rings, forming non-hexagonal lattices that differ markedly from the conventional honeycomb arrangement of \textit{h}-BN. In \textit{h}-BN, the hexagonal network is formed exclusively by alternating B--N nearest-neighbor bonds, as shown in Fig.~S1 of the Supporting Information. By contrast, the odd-membered rings present in PO-BN and PP-BN prevent complete B/N alternation. As a result, homonuclear B--B and N--N bonds are required to close the pentagonal rings. These bonds are therefore intrinsic to the proposed lattices and are essential for preserving their pentagon--octagon lattice geometry.

The PO-BN atomic network allows two possible distributions of the homonuclear B--B and N--N bonds. In one configuration, the N--N bond is located in the pentagonal rings while the B--B bonds occupy the octagonal rings; in the other, these positions are reversed. The former arrangement is energetically preferred by 347~meV/atom and is therefore adopted throughout this work (Fig.~\ref{fig1}). 
This pronounced energy difference indicates that the two homonuclear bonds are not energetically interchangeable within the pentagon–octagon network. Instead, the local geometric and electrostatic environments strongly influence their stability, with the weaker B–B bond being better accommodated by the less constrained octagonal rings, whereas placing it within the pentagonal rings incurs a substantially larger energetic penalty.
In contrast, the PP-BN lattice admits only a single bond arrangement because the homonuclear bonds lie on edges shared by pentagonal and octagonal rings. Consequently, interchanging B--B and N--N bonds does not produce a distinct configuration.

The optimized lattice parameters, characteristic bond lengths, and relative energies of both structures are summarized in Table~\ref{tab1}. PO-BN has equilibrium lattice constants of $a = 7.02$~\AA\ and $b = 5.14$~\AA, whereas PP-BN presents $a = 9.32$~\AA\ and $b = 3.80$~\AA, indicating that the two lattices accommodate the pentagon--octagon configuration through distinct in-plane arrangements.

\begin{table}[!htb]
\centering
\caption{Optimized lattice parameters, cell aspect ratio ($a/b$), B--N, B--B, and N--N bond-length ranges, and relative formation energy ($\Delta E$) of Pentaoctite-BN and Pop-BN monolayers.}
\begin{tabular}{l c c c c c c c}
\hline
Structure & $a$ (\AA) & $b$ (\AA) & $a/b$ & $d_{\mathrm{B-N}}$ (\AA) & $d_{\mathrm{B-B}}$ (\AA) & $d_{\mathrm{N-N}}$ (\AA) & $\Delta E$ (meV/atom) \\
\hline
PO-BN & 7.02 & 5.14 & 1.37 & 1.43--1.46 & 1.69 & 1.45 & 644 \\
PP-BN & 9.32 & 3.80 & 2.45 & 1.41--1.48 & 1.67 & 1.45 & 762 \\
\hline
\end{tabular}
\label{tab1}
\end{table}

In PO-BN, the B--N bond lengths range from 1.43 to 1.46~\AA, while the homonuclear B--B and N--N bonds are 1.69~\AA\ and 1.45~\AA. In PP-BN, the B--N bond lengths span a comparably narrow interval, from 1.41 to 1.48~\AA, while the B--B and N--N bonds are 1.67~\AA\ and 1.45~\AA, where both materials display a similarly regular local bonding environment. The pronounced structural distinction between the two polymorphs instead arises from the cell geometry and space-group symmetry, with PO-BN displaying a lower in-plane anisotropy, with a lattice-parameter ratio of $a/b \approx 1.37$ in the $P2/m$ space group, whereas PP-BN has a more elongated unit cell, with $a/b \approx 2.45$ in the lower-symmetry $Pm$ space group, nearly twice as anisotropic as PO-BN. This stronger geometric anisotropy of the unit cell suggests a more direction-dependent response in the elastic, vibrational, and electronic properties discussed below.

The energetic stability of PO-BN and PP-BN was quantified relative to hexagonal boron nitride (\textit{h}-BN), the experimentally realized
ground-state phase of 2D-dimensional BN~\cite{Cassabois2016,Novoselov2005}, through the relative energy per atom,

\begin{equation}
\Delta E = E_{\mathrm{BN\text{-}polymorph}}^{\mathrm{atom}} -
E_{h\text{-}\mathrm{BN}}^{\mathrm{atom}},
\label{eq:deltaE}
\end{equation}
where $E_{\mathrm{BN\text{-}polymorph}}^{\mathrm{atom}}$ and $E_{h\text{-}\mathrm{BN}}^{\mathrm{atom}}$ are the total energies per atom of the fully relaxed polymorph and of the \textit{h}-BN reference, respectively. The computed values are $\Delta E = 644$~meV/atom for PO-BN and 
$\Delta E = 762$~meV/atom for PP-BN, indicating that both polymorphs are metastable with respect to \textit{h}-BN. PO-BN is energetically favored over PP-BN by 118~meV/atom. As anticipated for binary lattices containing odd-membered rings, this energetic penalty is consistent with the presence of homonuclear B--B and N--N bonds imposed by the pentagonal rings of both geometries. The larger $\Delta E$ of PP-BN can be associated with its more distorted bonding environment, as reflected by the wider B--N bond-length distribution and the more anisotropic unit cell.

A positive $\Delta E$ does not, by itself, rule out possible experimental realization, provided that the phase is dynamically, thermally, and mechanically stable, as verified for both polymorphs in the following sections. The magnitude of $\Delta E$ can be placed in context by comparison with calculated relative energies reported for other non-hexagonal 2D lattices. Pentaoctite phases have been predicted for group-IV elements, with relative energies of 350, 120, 90, and 60~meV/atom for C, Si, Ge, and Sn, respectively, with respect to their corresponding hexagonal phases~\cite{Kegler2025}. Penta-graphene, another widely studied metastable carbon allotrope, is predicted to lie approximately 0.9~eV/atom above graphene~\cite{Zhang2015}. In addition, the biphenylene lattice, which has been synthesized by a bottom-up on-surface route~\cite{Fan2021}, lies 470~meV/atom above graphene in theoretical calculations~\cite{Kegler2025}. The values obtained for PO-BN and PP-BN are higher than those reported for elemental pentaoctite phases, but remain within the energy scale of metastable non-hexagonal 2D lattices discussed in the literature.

\begin{figure}[!htb]
    \centering
    \includegraphics[width=0.9\textwidth]{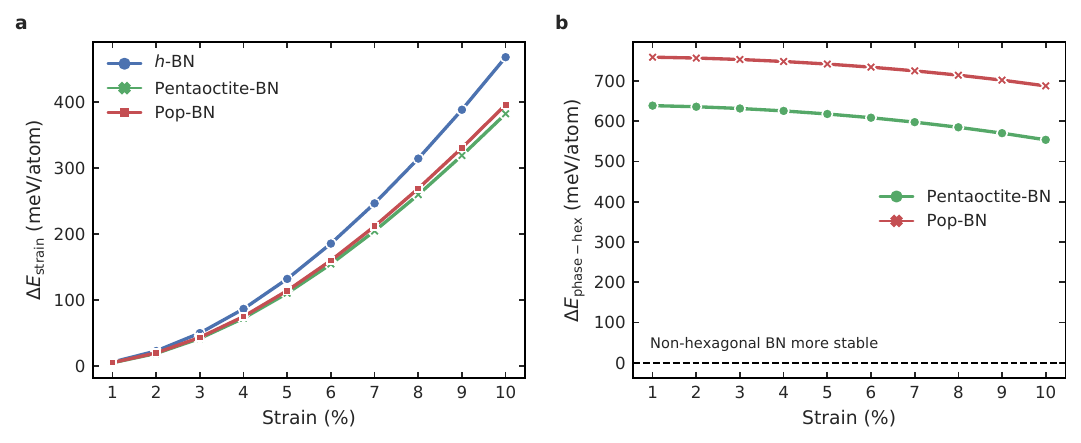}
    \caption{Strain-dependent energetics under biaxial tensile strain (1--10\%) for \textit{h}-BN, PO-BN, and PP-BN. (\textbf{a}) Strain energy per atom, $\Delta E_\mathrm{strain}$. (\textbf{b}) Relative stability of PO-BN and PP-BN with respect to \textit{h}-BN, $\Delta E_\mathrm{phase-hex}$; positive values indicate \textit{h}-BN remains more stable.}
\label{fig:strain_energetics}
    \label{fig_strain}
\end{figure}

To evaluate how the different lattice configurations accommodate mechanical deformation, we investigated the effect of biaxial tensile strain on the three BN phases. The strain-dependent energy provides a direct measure of the energetic penalty associated with lattice distortion and offers insight into the relative flexibility of the non-hexagonal networks compared to \textit{h}-BN. The response of the three phases to biaxial tensile strain was characterized by the strain energy per atom,
\begin{equation}
    \Delta E_{\mathrm{strain}}(\varepsilon) =
    \frac{E(\varepsilon) - E(0)}{N},
    \label{eq:strainE}
\end{equation}
where $E(\varepsilon)$ and $E(0)$ are the total energies of the strained and unstrained structures, respectively, $N$ is the number of atoms in the unit cell, and $\varepsilon$ is the biaxial strain referred to the equilibrium lattice parameters of each phase. As shown in Fig.~\ref{fig:strain_energetics}a, $\Delta E_{\mathrm{strain}}$ increases monotonically with strain for all three phases. Among the three structures, \textit{h}-BN exhibits the highest strain energy throughout the investigated range, reaching 470~meV/atom at 10\% strain, whereas PO-BN and PP-BN accumulate lower strain energies under the same deformation.

It is important to note that $\Delta E_{\mathrm{strain}}$ measures the total energy stored up to a finite strain and is therefore distinct from the linear elastic response. In particular, the lower strain energy of PO-BN does not necessarily imply a smaller Young's modulus, since its in-plane stiffness along the stiffest direction is comparable to that of \textit{h}-BN, as discussed in Sec.~\ref{sec:elastic}. Instead, the reduced finite-strain energy of PO-BN and PP-BN suggests a more compliant response in the nonlinear regime, likely associated with the ability of the pentagon--octagon networks to accommodate deformation through changes in bond angles and ring geometry, in addition to bond stretching. Similar strain-dependent energetic behavior has been reported for pentaoctite bismuth, where tensile strain drives a lattice geometry transition~\cite{Lima2019}, and for other group-V pentaoctites~\cite{daRosa2021}.

The evolution of the relative stability under strain was quantified by
\begin{equation}
    \Delta E_{\mathrm{phase\text{-}hex}}(\varepsilon) =
    \frac{E_{\mathrm{phase}}(\varepsilon)}{N_{\mathrm{phase}}} -
    \frac{E_{h\text{-}\mathrm{BN}}(\varepsilon)}{N_{h\text{-}\mathrm{BN}}},
    \label{eq:phasehex}
\end{equation}
where positive values indicate that the non-hexagonal phase is less stable than \textit{h}-BN at the same nominal biaxial strain $\varepsilon$. As shown in Fig.~\ref{fig_strain}b, $\Delta E_{\mathrm{phase}}$ remains positive over the full strain range, indicating that \textit{h}-BN remains the energetically preferred phase under tensile deformation. Starting from the zero-strain formation energies of 644 and 762~meV/atom for PO-BN and PP-BN, respectively (Table~\ref{tab1}), the energy separation decreases monotonically with strain, reaching 555 and 690~meV/atom at 10\% strain. These changes correspond to reductions of 89 and 72~meV/atom, respectively. Tensile strain therefore narrows the energetic separation between the metastable polymorphs and the hexagonal reference, although it does not invert the stability ordering within the range studied.

\begin{figure}[!h]
    \centering
    \includegraphics[width=0.8\textwidth]{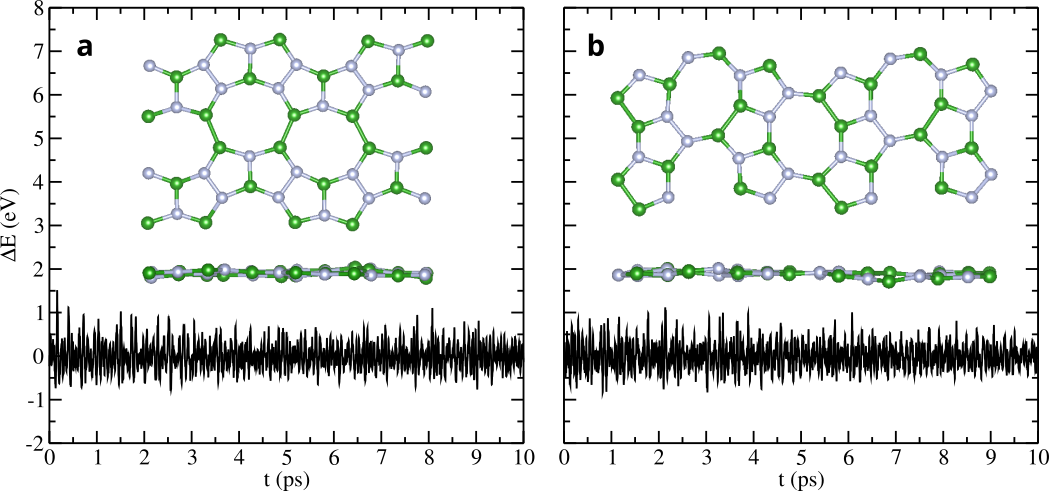} 
    \caption{Time evolution of the energy fluctuations, $\Delta E(t)$, during 
    10~ps AIMD simulations at 300 K for (a) PO-BN and (b) PP-BN monolayers. The final atomic configurations after the simulations are shown in top and side views.}
    \label{fig:aimd}
\end{figure}
The thermal stability of the proposed BN polymorphs was further evaluated by AIMD simulations performed at 300~K. Figure~\ref{fig:aimd} presents the time evolution of the energy fluctuations, defined as $\Delta E(t)=E(t)-\langle E\rangle$, together with the final atomic configuration after 10~ps of simulation. For PO-BN, Fig.~\ref{fig:aimd}(a), the energy fluctuates around its time-averaged value with no evidence of long-term drift or abrupt changes, indicating that the system remains in thermal equilibrium throughout the simulation.

Inspection of the final atomic configuration shows that the pentagon--octagon framework is preserved during the entire simulation. Although small deviations from the equilibrium atomic positions are observed, these distortions arise from thermal vibrations and do not involve bond breaking, lattice reconstruction, or any other irreversible structural transformation. 
A similar behavior is obtained for PP-BN, Fig.~\ref{fig:aimd}(b), confirming that both polymorphs remain structurally intact under ambient conditions.


\subsection{Electronic Properties}

\begin{figure}[!htb]
    \centering
    \includegraphics[width=0.8\textwidth]{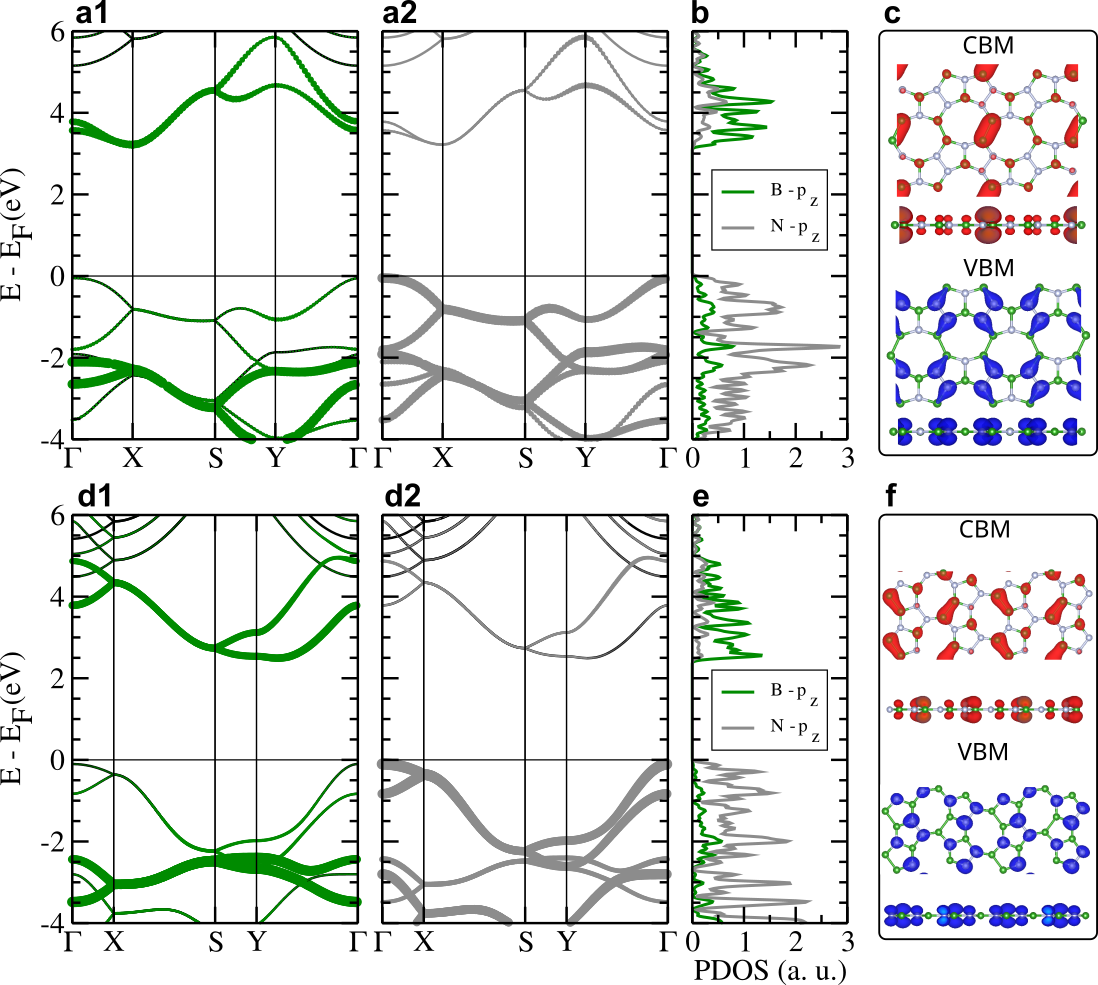} 
    \caption{Electronic properties of PO-BN (upper panels) and PP-BN (lower panels). (a1, d1) B-$p_z$ projected and (a2, d2) N-p$p_z$ projected electronic band structures; the size of the symbols represents the orbital contribution to each electronic state. (b, e) Projected density of states (PDOS) for the B-p$p_z$ and N-p$p_z$ orbitals. (c, f) Real-space charge-density distributions associated with the conduction-band minimum (CBM) and valence-band maximum (VBM), shown in top and side views.}
    \label{fig:bands}
\end{figure}

The electronic properties of PO-BN and PP-BN were investigated through their orbital-projected band structures, projected density of states (PDOS), and the real-space charge densities associated with the valence-band maximum (VBM) and conduction-band minimum (CBM), as shown in Fig.~\ref{fig:bands}. Both monolayers exhibit semiconducting behavior, although their band-edge dispersions reflect the distinct connectivity and symmetry of the two pentagon–octagon networks.

For PO-BN [Figs.~\ref{fig:bands}(a1) and \ref{fig:bands}(a2)], the VBM is located at the $\Gamma$-point, whereas the CBM is at the $X$-point, resulting in an indirect band gap of 3.26~eV. The bands near the Fermi level are dominated by the out-of-plane $p_z$ orbitals of B and N. The projected band structure shows that the upper valence states have predominantly N-$p_z$ character, whereas the lowest conduction band is mainly derived from B-$p_z$ states. 
This orbital separation is also evident in the PDOS plot [Fig. 4(b)], where the states immediately below the Fermi level are dominated by N-$p_z$ contributions, while the conduction-band edge is mainly associated with B-$p_z$ orbitals. 
This behavior is characteristic of the polar nature of B–N bonding, in which the higher electronegativity of N stabilizes the occupied states, whereas the low-lying unoccupied states are preferentially associated with B atoms.
The spatial distributions of the band-edge states [Fig.~\ref{fig:bands}(c)] provide a real-space representation of this orbital character. The VBM charge density is predominantly localized around the N sublattice and displays lobes extending perpendicular to the monolayer plane, consistent with its N-$p_z$ character. In contrast, the CBM density is mainly distributed around the B sites, also with a pronounced out-of-plane character.

PP-BN exhibits a similar orbital organization but a distinct band dispersion [Figs.~\ref{fig:bands}(d1) and \ref{fig:bands}(d2)]. The VBM is located at $\Gamma$, whereas the conduction-band minimum occurs along the $Y-\Gamma$ path,
also indicating an indirect-gap semiconductor, but with a smaller band gap of 2.59~eV. As in PO-BN, the states close to the band edges are dominated by $p_z$ orbitals. The upper valence bands are primarily associated with N-$p_z$ states, while the lowest conduction band has a dominant B-$p_z$ contribution. The PDOS shown in Fig.~\ref{fig:bands}(e) confirms this complementary distribution of the band-edge states. Compared with PO-BN, the different dispersion of the frontier bands reflects the more anisotropic PP-BN lattice and its distinct arrangement of pentagonal and octagonal rings.
The VBM and CBM charge densities of PP-BN [Fig.\ref{fig:bands}(f)] corroborate the projected band analysis. The VBM is primarily distributed over N atoms, whereas the CBM is concentrated mainly around the B sublattice. Both states display the characteristic out-of-plane lobes expected from $p_z$-derived states.
\subsection{Elastic Properties}
\label{sec:elastic}

The elastic stability and in-plane mechanical response of PO-BN and PP-BN 
monolayers were evaluated from the elastic constants $C_{11}$, $C_{22}$, 
$C_{12}$, and $C_{66}$, as well as the Young's modulus ($E$) and Poisson's 
ratio ($\nu$). The calculated values are summarized in 
Table~\ref{tab:elastic}. For the investigated 2D systems, the Born-Huang 
stability criteria require $C_{11}>0$, $C_{22}>0$, $C_{66}>0$, and 
$C_{11}C_{22}>C_{12}^{2}$ \cite{Born1940,born1954dynamical}. The positive elastic constants and 
fulfillment of these conditions confirm that both PO-BN and PP-BN are 
mechanically stable against small in-plane deformations.

\begin{table}[!htb]
\centering
\caption{Calculated elastic constants, Young's modulus range, and 
Poisson's ratio range of Pentaoctite-BN and Pop-BN monolayers.}
\label{tab:elastic}
\resizebox{\textwidth}{!}{
\begin{tabular}{lcccccccc}
\hline
Structure 
& $C_{11}$ (N/m) 
& $C_{22}$ (N/m) 
& $C_{12}$ (N/m) 
& $C_{66}$ (N/m) 
& $E_{\min}$ (N/m) 
& $E_{\max}$ (N/m) 
& $\nu_{\min}$ 
& $\nu_{\max}$ \\
\hline

Pentaoctite-BN  
& 290.44 & 234.35 & 43.10 & 113.44 
& 227.96 & 282.52 
& 0.14 & 0.18 \\

Pop-BN  
& 221.24 & 262.59 & 49.88 & 74.82  
& 195.46 & 251.34 
& 0.19 & 0.32 \\

\hline
\end{tabular}}
\end{table}

The directional dependence of the mechanical properties was further analyzed 
through the angular variation of the Young's modulus $E(\theta)$ and 
Poisson's ratio $\nu(\theta)$, obtained from the elastic compliance tensor  \cite{Cadelano2010}:
\begin{equation}
E(\theta)=
\frac{X}
{C_{11}a^{2}+\left(\frac{X}{C_{66}}-2C_{12}\right)ab+C_{22}b^{2}},
\end{equation}
\begin{equation}
\nu(\theta)=
\frac{C_{12}a^{2}-\left(C_{11}+C_{22}-\frac{X}{C_{66}}\right)ab+C_{12}b^{2}}
{C_{11}a^{2}+\left(\frac{X}{C_{66}}-2C_{12}\right)ab+C_{22}b^{2}},
\end{equation}
where $a=\sin^{2}\theta$, $b=\cos^{2}\theta$, and $X=C_{11}C_{22}-C_{12}^{2}$.

\begin{figure}[!htb]
    \centering
    \includegraphics[width=0.7\columnwidth]{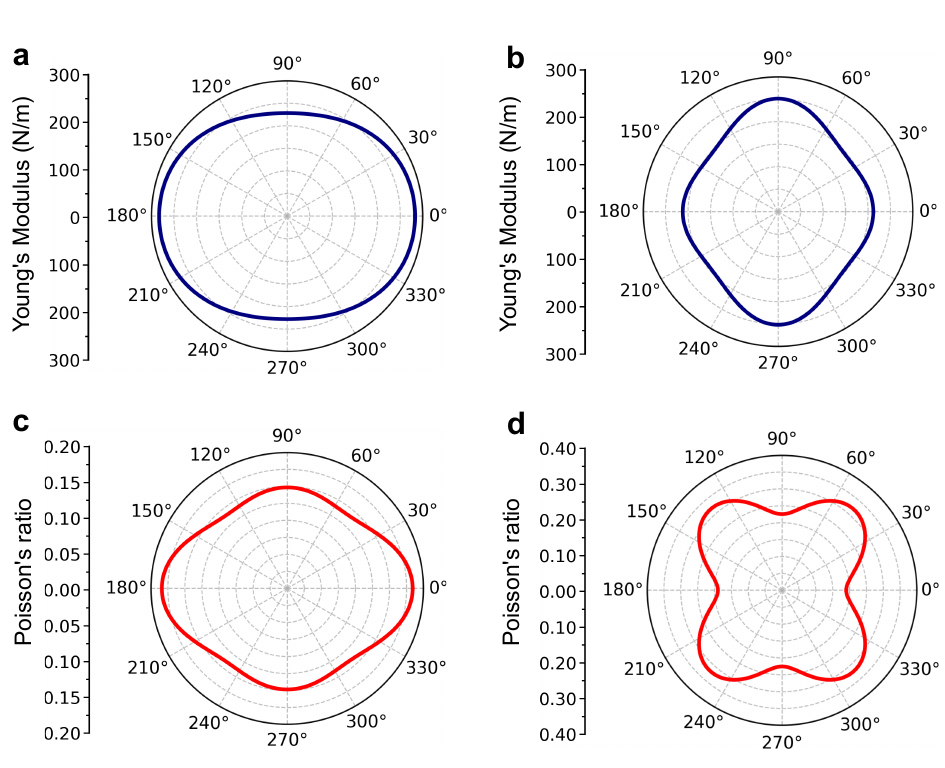}
    \caption{Angular dependence of the Young's modulus, $E(\theta)$, and Poisson's ratio, $\nu(\theta)$, for Pentaoctite-BN [(a), (c)] and Pop-BN [(b), (d)]. The Young's modulus is expressed in N/m, whereas the Poisson's ratio is dimensionless.}
    \label{fig:elastic}
\end{figure}

The two monolayers display markedly different anisotropy patterns (Fig.~\ref{fig:elastic}, Table~\ref{tab:elastic}), which can be traced to their distinct ring arrangements (Fig.~\ref{fig1}). Both structures are strictly planar, and in both the B--N bonds are nearly homogeneous in length ($1.43$--$1.47$~\AA\ for PO-BN and $1.40$--$1.46$~\AA\ for PP-BN; Table~\ref{tab1}), so the difference in elastic anisotropy does not originate from bond-length disparities. Instead, it reflects how each network accommodates strain through bond-angle distortion. In PO-BN, the octagons and pentagons are aligned with the cell axes, the unit cell is only moderately elongated ($a/b=1.37$), and the Young's modulus [Fig.~\ref{fig:elastic}(a)] is smooth and nearly elliptical, with the stiff and soft directions coinciding with the principal axes ($E_{\max}/E_{\min}=1.24$). In PP-BN, the pentagon--octagon motifs are tilted with respect to the cell axes and the unit cell is strongly elongated ($a/b=2.45$); the Young's modulus [Fig.~\ref{fig:elastic}(b)] develops a four-lobed profile with extrema rotated away from the principal axes and a larger anisotropy ($E_{\max}/E_{\min}=1.29$). Since the bond lengths are comparable in both phases, this stronger and rotated anisotropy is attributed to the angular (bond-bending) stiffness of the tilted ring arrangement rather than to bond stretching. 

Consistent with this picture, PP-BN deviates more strongly from the elastic isotropy condition $2C_{66}=C_{11}+C_{22}-2C_{12}$ than PO-BN (61\% versus 48\% relative deviation), reflecting its lower shear modulus ($C_{66}=74.82$~N/m, against $113.44$~N/m for PO-BN; Table~\ref{tab:elastic}). This angular softening is independently corroborated by the phonon analysis in Section~\ref{sec:phonons}: PP-BN's scissor-type bending modes extend to lower frequency than PO-BN's ($241$ versus $359$~cm$^{-1}$; Table~\ref{tab:phonon}). The same angular softening is independently corroborated by the phonon analysis presented in Section~\ref{sec:phonons}, where PP-BN exhibits scissor-type bending modes extending to substantially lower frequency than in PO-BN (down to $241$~cm$^{-1}$, against $359$~cm$^{-1}$; Table~\ref{tab:phonon}), consistent with a softer angular restoring force in the tilted ring network.

Poisson's ratio mirrors this behavior. For PO-BN [Fig.~\ref{fig:elastic}(c)], $\nu(\theta)$ varies between $0.14$ and $0.18$, close to the value reported for \textit{h}-BN ($\nu\approx0.21$) \cite{Falin2017}. For PP-BN [Fig.~\ref{fig:elastic}(d)], $\nu(\theta)$ spans a wider range, from $0.19$ up to $0.32$ along the compliant directions, indicating a stronger transverse coupling than in either PO-BN or \textit{h}-BN. In both monolayers $\nu(\theta)$ remains positive over the full angular range, so neither structure is auxetic.

\subsection{Phonon Dispersion and Vibrational Mode Analysis}
\label{sec:phonons}

\begin{figure}[!htb]
    \centering
    \includegraphics[width=0.8\textwidth]{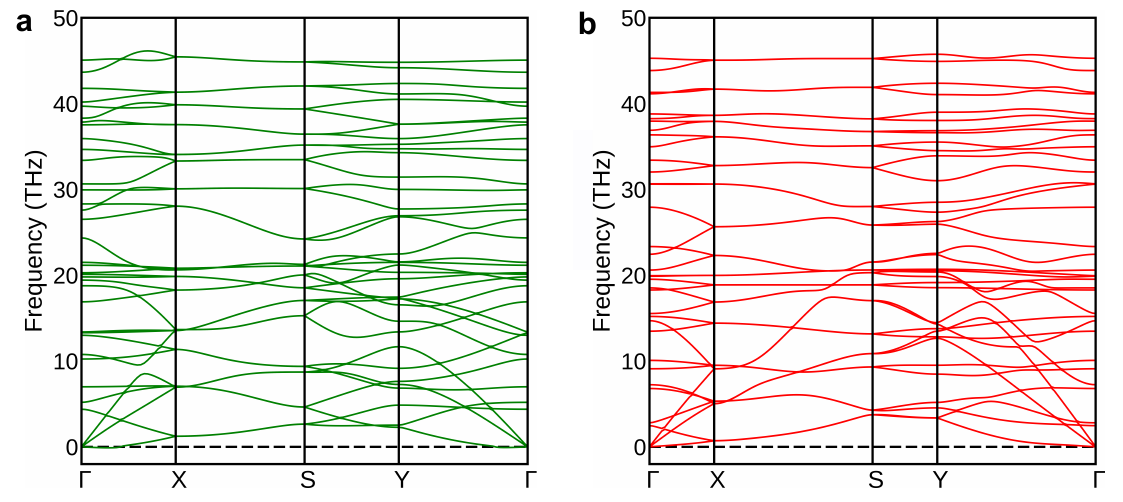} 
    \caption{Phonon dispersion relations of (a) PO-BN and (b) PP-BN calculated along the high-symmetry path of their respective Brillouin zones.}
    \label{fig:phonon}
\end{figure}

Building on the mechanical stability established above, the dynamical stability of the proposed BN polymorphs was further evaluated through their phonon dispersion relations, shown in Fig.~\ref{fig:phonon}. In both PO-BN and PP-BN, all phonon branches remain non-negative throughout the entire Brillouin zone, indicating the absence of soft vibrational modes that would drive spontaneous structural distortions. Together with the elastic constants satisfying the Born-Huang criteria, this confirms that both lattices correspond to dynamically robust local minima on the potential-energy surface, despite their relatively higher energies with respect to \textit{h}-BN.

The phonon spectra also reflect the distinct structural characteristics of the two polymorphs. PO-BN exhibits a more uniform distribution of phonon branches, consistent with its higher crystallographic symmetry and narrower bond-length distribution. In contrast, PP-BN displays a more complex phonon spectrum with a larger dispersion of the optical branches, reflecting its lower symmetry and more anisotropic bonding environment.

For both structures, the acoustic branches exhibit the expected behavior of two in-plane modes with approximately linear dispersion near the $\Gamma$ point and one flexural out-of-plane (ZA) mode with quadratic dispersion, as required for stable free-standing 2D crystals. The absence of imaginary frequencies, together with the favorable energetic response under biaxial strain discussed above, provides strong evidence that the proposed PO-BN and PP-BN monolayers are not only energetically metastable but also dynamically robust, supporting their potential experimental realization.

To further characterize the optical phonon manifold, each $\Gamma$-point mode was classified according to the relative in-plane displacement of bonded B--N pairs, modes with displacement predominantly parallel to the B--N bond are identified as bond-stretching modes, while modes with displacement predominantly perpendicular to the bond, within the layer plane, are identified as scissor (bending) modes. Out-of-plane modes were separated by their dominant displacement component along $z$. The resulting frequency ranges are summarized in Table~\ref{tab:phonon}.

\begin{table}[!htb]
\centering
\caption{Frequency ranges of stretching and scissor optical phonon modes at $\Gamma$, and the highest-frequency stretching and most polarized scissor mode, for PO-BN and PP-BN monolayers. All frequencies are given in cm$^{-1}$.}
\label{tab:phonon}
\resizebox{\textwidth}{!}{
\begin{tabular}{lcccccc}
\hline
Structure
& Stretching range
& Scissor range
& $n_{\mathrm{stretch}}$
& $n_{\mathrm{scissor}}$
& Highest stretching mode
& Most polarized scissor mode \\
\hline
Pentaoctite-BN
& 812--1503 & 359--1198 & 6 & 8
& 1502.9 (93\% stretching)
& 563.5 (98\% scissor) \\
Pop-BN
& 1068--1510 & 241--1022 & 5 & 9
& 1510.1 (85\% stretching)
& 778.4 ($>$99\% scissor) \\
\hline
\end{tabular}}
\end{table}

In both monolayers, the highest-frequency optical branch corresponds to a nearly pure B--N bond-stretching mode (1502.9~cm$^{-1}$ for PO-BN, 1510.1~cm$^{-1}$ for PP-BN), with comparable stretching character (93\% and 85\%, respectively) and nearly identical frequency, indicating very similar intrinsic stiffness of the shortest B–N bonds in the two polymorphs.

The two structures differ markedly in the scissor-mode manifold. In PO-BN, scissor character is confined to a narrower window (359--1198~cm$^{-1}$), with the most polarized scissor mode found at 563.5~cm$^{-1}$ (98\% scissor character). In PP-BN, scissor modes extend to substantially lower frequency (241--1022~cm$^{-1}$), with an essentially pure scissor mode at 778.4~cm$^{-1}$. The broader and more fragmented stretching manifold of PO-BN (six modes spanning 812--1503~cm$^{-1}$, versus five modes spanning 1068--1510~cm$^{-1}$ in PP-BN) mirrors the narrower B--N bond-length distribution previously discussed (Table~\ref{tab1}), while the wider scissor window of PP-BN reflects its larger structural anisotropy ($a/b \approx 2.45$) and lower space-group symmetry ($Pm$).

The calculated $\Gamma$-point frequencies and mode characters provide reference values for the future experimental identification of PO-BN and PP-BN by Raman and infrared spectroscopy. In particular, the near-degenerate, high-frequency B--N stretching modes (1502.9~cm$^{-1}$ and 1510.1~cm$^{-1}$ for PO-BN and PP-BN, respectively) fall close to the well-characterized $E_{2g}$ stretching mode of \textit{h}-BN ($\sim$1366~cm$^{-1}$), offering a practical spectral marker to distinguish these polymorphs from both \textit{h}-BN and from each other once synthesized.

\subsection{Optical response}

The optical response of the three BN polymorphs reveals that lattice geometry engineering provides an efficient route to tune the position of the excitonic series over a broad spectral range while maintaining the strong excitonic character typical of 2D BN materials.

\begin{figure}[!htb]
    \centering
    \includegraphics[width=0.85\textwidth]{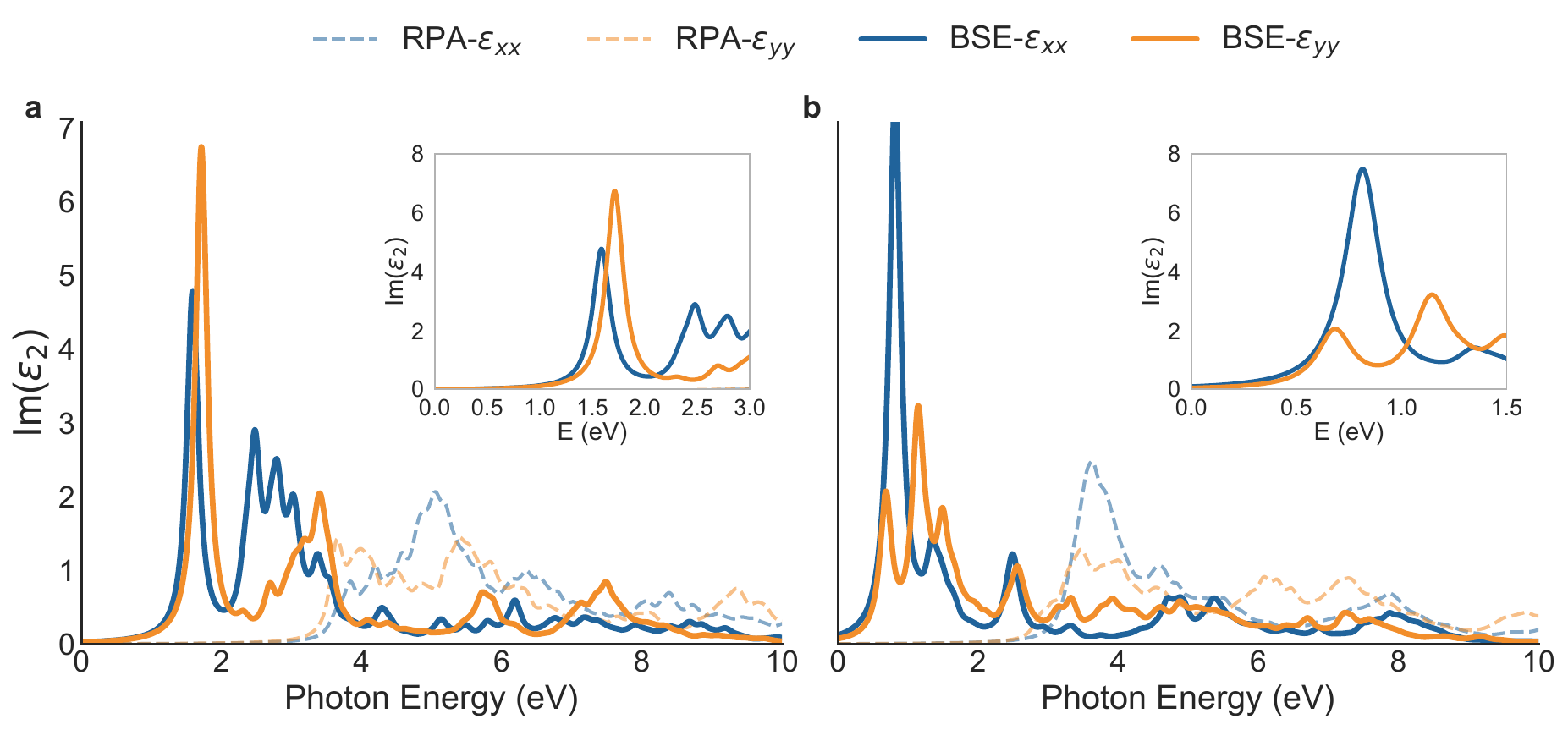} 
    \caption{Imaginary part of the in-plane dielectric function of (a) PO-BN and (b) POP-BN obtained within the RPA (dashed lines) and BSE (solid lines)
frameworks using the HSE06 electronic-structure reference. Blue and orange curves correspond to the $\varepsilon_{xx}$ and $\varepsilon_{yy}$ components, respectively. Insets highlight the excitonic absorption onset.}
    \label{fig:optical_response}
\end{figure}

Figure~\ref{fig:optical_response} compares the in-plane dielectric functions obtained within the RPA and BSE formalisms for PO-BN and PP-BN, while the corresponding spectra for monolayer \textit{h}-BN are presented in the Fig.~S2 of the Supporting Information. In all three systems, the inclusion of electron-hole interactions via the BSE formalism yields a pronounced redshift of the absorption onset relative to the independent-particle RPA spectra, indicating the strong Coulomb interaction and weak dielectric screening characteristic of atomically thin BN materials. In contrast, the out-of-plane response ($\varepsilon_{zz}$) remains essentially dark throughout the excitonic region due to the optical selection rules imposed by the planar BN lattice.

The first intense peaks emerging below the RPA continuum correspond to bound bright excitons, indicating that the optical gap is substantially smaller than the HSE06 electronic gap. In PO-BN, the absorption onset is defined by an excitonic doublet at 1.588~eV ($\varepsilon_{xx}$) and 1.712~eV ($\varepsilon_{yy}$), resulting in an anisotropy splitting of 124~meV. PP-BN exhibits an even stronger redshift, with the first bright exciton appearing at 0.684~eV in the $\varepsilon_{yy}$ channel, followed by the dominant $x$-polarized excitation at 0.816~eV. Remarkably, the anisotropy splitting remains nearly unchanged at 132~meV, although the polarization ordering is reversed relative to PO-BN.

A clear hierarchy emerges across the three polymorphs. While the energy separation between the first bright exciton and the onset of the independent-particle continuum remains close to 2~eV in all cases, the excitonic series shifts dramatically toward lower energies following the reduction of the electronic gap induced by lattice engineering. Consequently, \textit{h}-BN, PO-BN, and PP-BN progressively extend the optical activity from the ultraviolet to the visible and finally to the telecommunication infrared region.

The coexistence of strong excitonic effects and pronounced in-plane anisotropy makes PO-BN and PP-BN promising candidates for polarization-sensitive photodetectors and optical modulators. In particular, the lowest bright exciton of PP-BN at 0.684~eV corresponds to approximately 1813~nm in the short-wavelength infrared region, while the dominant excitation at 0.816~eV lies close to the 1550~nm telecommunication window. Together with the ultraviolet activity of \textit{h}-BN, these results demonstrate that lattice engineering alone enables a single BN chemistry to span deep ultraviolet, visible, near-infrared, and optical communication wavelength ranges.

\section*{Conclusions}

In summary, we have proposed two novel non-hexagonal boron nitride monolayers, PO-BN and PP-BN, and systematically investigated their structural, electronic, mechanical, vibrational, thermal, and optical properties using first-principles calculations. Although metastable with respect to hexagonal BN, both polymorphs satisfy the criteria for dynamical, mechanical, and thermal stability, demonstrating that they are viable two-dimensional materials.
Both monolayers are indirect-gap semiconductors whose electronic states near the band edges are dominated by out-of-plane $p_z$ orbitals. Their distinct pentagon–octagon ring networks give rise to markedly different elastic anisotropies while preserving the strong excitonic effects characteristic of atomically thin BN. Most importantly, lattice engineering shifts the optical absorption onset from the deep-ultraviolet region of h-BN to the visible range in PO-BN and to the telecommunication infrared in PP-BN.

These findings demonstrate that tailoring the lattice architecture provides an effective route to tailor the electronic and optical response of two-dimensional BN without changing its chemical composition, establishing PO-BN and PP-BN as promising platforms for polarization-sensitive optoelectronic and photonic devices.

\section*{Acknowledgments}

The authors acknowledge the financial support provided by FAPEMIG. They also gratefully acknowledge the computational resources provided by LCC-UFLA and CENAPAD-SP. E.N.L. acknowledges the support of INCT Materials and INCT Photonics.





\bibliographystyle{iopart-num}
\bibliography{reference}

\end{document}